# Injection-locked phonon lasing driven by an individual plasmonic metamolecule

Brian J. Roxworthy[1,*] and Vladimir A. Aksyuk[1]

[1] Center for Nanoscale Science and Technology, National institute of Standards and Technology, 100 Bureau Drive, Gaithersburg, MD 20899, USA
[*] To whom correspondence should be addressed: brian.roxworthy@nist.gov

Nanophotonic devices take advantage of geometry-dependent optical properties to confine and enhance the interaction of light with matter on small scales.  By carefully patterning nanoscale geometries, coupling of responses across distinct physical domains can be achieved; such as photon and phonon modes in optomechanical crystals.  Extreme optical confinement of plasmonics can localize strong cross-domain interactions into nanometer-scale volumes, smaller than many nanosystem's thermal and mechanical eigenmodes, opening new regimes for engineering reconfigurable, dynamic nanosystems. Here, we present a nanosystem based on an electrically actuated, mechanically coupled gap-plasmon metamolecule resonator.  Integrated electrical actuation reveals distinct dispersive and reactive optomechanical coupling regimes of the metamolecules, and enables both broad spectral tuning of individual plasmonic resonances and selective optical transduction of nanomechanical modes. Strong thermomechanical backaction from a single point-like metamolecule yields optically driven self-oscillation, and we show that this metamolecule phonon laser can be injection-locked to enhance weak mechanical stimuli. This multi-domain nanosystem advances nanomechanical sensing and optical modulation, opens avenues for studying nanoscale nonlinear optomechanics, and may serve as a building block for future "smart" metamaterials.

Size effects are the hallmark of nanoscale systems: strong dependence of physical responses on the shape of structural elements creates, for example, the optical resonances of dielectric or noble-metal plasmonic nanostructures  [1,2], and gives rise to many unusual properties of metamaterials. Careful structuring of individual building blocks tailors their physical responses and enables remarkable phenomena such as cloaking of light, sound, or water waves  [3], super resolution imaging  [4], or diode-like mechanical behavior  [5]. Both energy confinement in the structure and geometric dependencies can strongly enhance interactions between modes originating from multiple physical domains within a single nanoscale system.  This cross-domain coupling is leveraged in cavity optomechanics and nanoelectromechanical (NEMS) systems  [6,7], and opens a path toward technologically important applications.  However, harnessing multi-domain interactions to their full potential requires engineering strongly interacting modes in nanometer-localized volumes.

Plasmomechanical systems are an attractive candidate to this end.  They are conceptually similar to cavity optomechanical devices, but operate in a complementary regime characterized by broadband optical performance, extremely strong optomechanical coupling, and deep subwavelength mode volumes  [8–10].  Arrays of electrically and optically actuated plasmomechanical systems hold great promise for shaping and controlling optical fields, and to date there has been progress showing extended surfaces with tunable optical response  [11].  The optical power absorbed by plasmonic structures can be harnessed to produce a large mechanical response to light through purposeful coupling of thermal and mechanical domains.  While this thermomechanical backaction has been demonstrated with large-area plasmonic arrays supporting modes many wavelengths across in two dimensions  [12], creating individually tunable subwavelength plasmonic metamolecules and harnessing single metamolecule backaction on a nanomechanical system has not been demonstrated.  Such nanoscale systems can not only serve as enhanced, miniaturized sensors and transducers, but also form building blocks for future active 'smart' metamaterials and metasurfaces akin to smart materials with engineered coupled responses in multiple physical domains.

In this article, we demonstrate a multi-domain nanosystem based on an electrically tunable, subwavelength plasmonic metamolecule with engineered strong coupling between optical, thermal, and mechanical modes. Leveraging multiple interaction pathways between these modes, we show that the strength and the spectral properties of optical coupling to thermal and mechanical modes can be flexibly engineered via the spatial location of the point-like metamolecule and controlled with electrical actuation. This capability leads to broad tuning of individual localized plasmonic resonances, and self-oscillation and cooling of the nanomechanical system *via* strong thermomechanical backaction of the individual subwavelength metamolecule. Further, we demonstrate electromechanical injection locking of this metamolecule phonon laser, which can function as an amplifier for sensitive readout of a weak periodic signal frequency.

Figure 1a illustrates the geometry of our devices. Each device in this work comprised a silicon nitride cantilever with approximately 2 µm width and a length of either 4 µm or 6 µm with a gold cuboid nominally measuring (350×160×40) nm$^3$ embedded on the underside at varying positions along the cantilever length (Figs. 1d,e). The cantilever can be electrostatically actuated by applying a voltage between a thin (≈ 15 nm) gold electrode placed on its top surface and an underlying stationary gold pad. Chips containing arrays of complete devices were packaged and wirebonded on a printed circuit board (Fig. 1b). A residual stress gradient caused the cantilevers to curl away from the substrate, increasing the nominal 20 nm gap at the base quadratically along the cantilever length (Fig. 1c); full fabrication details are given in the Methods and Supplementary Fig. 1.

The optical metamolecule comprises the movable nanoscale gold cuboid and stationary pad separated by a gap ($g$); we define the motion coordinate $x$ as positive toward smaller gap. The metamolecule is designed with an $m = 3$ localized gap plasmon (LGP) optical frequency resonance, wherein the round-trip phase of the Fabry-Perot like standing wave gap plasmon is $2\pi m$ [9,13]. As shown in the calculated reflectance ($R$) curves (Fig. 2a, Supplementary Fig. 2), this LGP mode manifests as a pronounced Lorentzian-shaped dip centered at wavelength (frequency) $\lambda_{\mathrm{LGP}}$ ($\omega_{\mathrm{LGP}} = 2\pi c/\lambda_{\mathrm{LGP}}$, $c$ being the vacuum speed of light). A unique feature of this optomechanical metamolecule is that it exhibits two regimes of optomechanical coupling: reactive and dispersive.

When positioned further than approximately 1.5 µm from the base (Figs. 1c,e) the gaps, $g > 35$ nm, are large and $\omega_{\mathrm{LGP}}$ is nearly independent of $x$. In this case, the dispersive optomechanical coupling constant $g_{\mathrm{om}} = -\partial \omega_{\mathrm{LGP}}/\partial x \approx 0$ (Figs. 2a,b). However, in this regime the LGP mode's coupling to radiation increases with larger gap, and therefore the metamolecule responds primarily by increasing its coupling depth (less light reflected on resonance) with increasing gap. The result is a reactive optomechanical coupling [14], characterized by $\partial A/\partial x < 0$ for input wavelengths, where $A \approx 1 - R$ is the absorbance. Conversely, the smaller-gap metamolecules ($g < 35$ nm) closer to the cantilever base (Fig. 1d) acquire a dispersive coupling component that shifts $\lambda_{\mathrm{LGP}}$ ($\omega_{\mathrm{LGP}}$) to longer wavelengths (lower frequencies) as the gap decreases. The dispersive $g_{\mathrm{om}}/2\pi$ can reach exceptionally large values of 7 THz nm$^{-1}$ (Figs. 2a,b).

When an actuator voltage is applied, the electrostatic force decreases the gap and strongly tunes the metamolecule optical properties. We take advantage of this actuation to show strong optical modulation of individual metamolecule resonances and, through plasmomechanical motion transduction, direct measurement of the different optomechanical coupling regimes. Figure 2c shows experimentally measured reflectance spectra for increasing applied voltages with a metamolecule located ≈ 1.5 µm from the base of a ≈ 4 µm cantilever, measured using a confocal spectroscopy setup with a broadband supercontinuum laser (Methods, Supplementary Fig. 3). We measured shifts in the

LGP resonance $\lambda_{\text{LGP}}$ = (763 ± 0.1) nm up to a maximum of $\delta\lambda_{\text{LGP}}$ = (42 ± 1.8) nm, with the native linewidth $\Delta\lambda$ = (39 ± 0.5) nm (Fig. 2d); uncertainties are derived from Lorentzian fits. These values give a relative shift of $\delta\lambda_{\text{LGP}}/\Delta\lambda$ = 1.07 ± 0.05 (Fig. 2e). For a similar device having a metamolecule closer to the cantilever base (≈ 0.5 µm), the same voltage induces a smaller motion and thus a smaller shift $\delta\lambda_{\text{LGP}}/\Delta\lambda$ = 0.35 ± 0.04. Achieving $\delta\lambda_{\text{LGP}}/\Delta\lambda$ > 1 at a CMOS (complementary metal-oxide semiconductor) compatible voltage of 2.75 V means that the phase of the light re-radiated by the device is electromechanically tuned by more than $\pi$ rad, representing a technologically important milestone for implementing efficient phase and high-contrast amplitude modulators. Indeed, the measured intensity modulation of up to 40 % (Fig. 2c, inset) shows that a single tunable metamolecule can act as an effective far-field optical modulator, notwithstanding its sub-wavelength size. Along these lines, we find that our devices have an effective electro-optic Kerr coefficient of approximately 2×10$^{-8}$ m V$^{-2}$, which is extremely large compared to ordinary Kerr media (see Supplementary Information) [15,16].

The reactive and dispersive optomechanical couplings can be distinguished by measuring the optical response to electrostatic actuation. We determined the coupling by applying a small ≈ 40 mV voltage oscillating at varying input frequency to excite cantilever vibrations, while the motion was read out *via* metamolecule optomechanical transduction [9] (Methods, Supplementary Fig. 3). Figures 2f,g show the magnitude and phase responses, respectively, of two separate 4 µm cantilevers having metamolecules at ≈ 0.5 µm and ≈ 2.0 µm from the cantilever base. In both cases, the same fundamental and second order flexural modes near 10 MHz and 55 MHz, respectively, are visible in the magnitude response. For the 0.5 µm device, a significant motion signal is present only for nonzero wavelength detuning ($\delta\lambda \neq 0$). Further, a clear reversal of the phase response sign is evident when switching from a blue to a red-detuned probe, the latter being shifted to $\pi$-out of phase with the motion. These observations are consistent with dispersive coupling. In contrast, the 2.0 µm device transduces motion regardless of $\delta\lambda$, with stronger normalized magnitude on resonance ($\delta\lambda \approx 0$), and the phase identical for all detunings. This behavior is consistent with reactive coupling [14].

Choosing different locations for the point-like metamolecule within the cantilever controls the strength of its optical interaction with different mechanical modes: while both cantilevers have the same modes, for the 0.5 µm metamolecule location, the second order mode dominates the magnitude response, whereas the fundamental mode dominates at the 2.0 µm location. We achieve similar transduction of electrically stimulated motion using much smaller cantilevers measuring approximately 500 nm × 2 µm and having an embedded ≈ (75 × 90) nm metamolecule (Supplementary Fig. 4). This metamolecule supports an $m$ = 1 LGP mode [9], and is much smaller than the diffraction limit of our optical system. Consequently, these results show that the dynamic response of this electrically stimulated nanomechanical system is measured locally, at the position of the probe metamolecule.

Plasmonic absorption generates heat, which locally couples the optical mode to the thermal modes of the structure. In turn, the fundamental thermal mode is deliberately coupled to the fundamental mechanical mode via thermal expansion mismatch between the top metal electrode and the nitride cantilever. This strong thermomechanical coupling enriches the system's behavior, allowing an individual metamolecule to drive optically-powered self-oscillation of a nanomechanical system, forming a metamolecule phonon laser. To demonstrate this thermomechanical backaction, we placed our devices into a vacuum chamber, primarily to reduce the effects of squeeze-film damping from ambient air. Here, we use a longer 6 µm cantilever with a dispersively coupled metamolecule located ≈ 0.5 µm from the base and focus a wavelength-tunable pump laser onto the device (Methods, Supplementary Fig. 3). This longer device has a lower mechanical frequency of ≈ 4 MHz. Figure 3a shows the motion spectra of the cantilever's fundamental mode, transduced by the metamolecule with a blue-detuned laser; calibration was performed using equipartition at low laser power [17,18] (Methods, Supplementary Information). The measured motion amplitude (Fig. 3b) increases gradually up to a pump power of approximately 1100 µW, beyond which a sharp transition occurs, whereby the amplitude increases by more than 50 dB and saturates at high power. Associated with this transition is a reduction of the measured linewidth (Fig. 3b inset). The combination of a distinct threshold and reduction in mechanical linewidth indicates the device operates as a phonon laser, with the feature that the stimulated emission of coherent phonons occurs without a population inversion [19–23]. Such behavior has been observed in single-ion systems [23] and is analogous to stimulated Raman scattering in optical Raman lasers [24]. Control experiments have verified that excitation of the LGP resonance is required both to transduce motion and to observe phonon lasing (Supplementary Fig. 5). In contrast, increasing the power of a red-detuned pump dampens mechanical motion (Fig. 3b) and broadens the mechanical linewidth (Fig. 3b, inset), in agreement with expectations for a dispersively coupled device [7]. The blue-detuned narrowing is stronger than red-detuned broadening, which is expected because coupling is not purely dispersive (Fig. 3c). Our system has the unique feature that the optically pumped phonon lasing is driven by a single point-like metamolecule, and therefore the lasing dynamics can be tuned by changing the metamolecule position or geometry. Indeed, for reactively coupled devices, lasing is observed for both positive and negative wavelength detuning (Fig. 3c).

Phonon lasing occurs due to time-delayed thermal feedback between the gap-size dependent absorption in the metamolecule and thermal bimorph actuation of the cantilever. For a given temperature increase $\Delta T$, the bimorph closes the gap with gain $g_B \equiv \partial x/\partial \Delta T \approx 45$ pm K$^{-1}$ at the location of the metamolecule (Supplementary Fig. 2). The finite time constant $\tau_t \approx 1.5$ µs of the thermal mode's response causes a phase-delay between motion and bimorph actuation, resulting in a complex-valued effective photothermal spring that applies a force component proportional to velocity. For blue-detuned pumping of the dispersive metamolecule, the absorbance gain $g_A$ satisfies $g_A \equiv \partial A/\partial x < 0$ and the photothermal spring contributes negative damping to the system, adding energy to the mechanical oscillation each cycle (Fig. 3d). Lasing occurs when the overall photothermal gain exceeds intrinsic mechanical losses, occurring for a threshold pump power

$$P_0 \geq \frac{m_{th} \, c_p}{|g_A|} \frac{\omega_m^2 + \tau_t^{-2}}{Q_m \, \omega_m \, g_B}, \tag{1}$$

where $m_{th} \approx 4.6$ pg is the cantilever thermal mass, $Q_m$ = 1655 $\pm$ 35 is the native mechanical quality factor, and $\omega_m = 2\pi \cdot 4030$ kHz is the resonant mechanical frequency; a mass-specific heat of $c_p$ = 1100 J kg$^{-1}$ K$^{-1}$ is assumed (see Supplementary Information). Figure 3c shows the numerically calculated absorbance gain landscape for our system, with the blue and red-detuned measurement conditions indicated as white and black data points, respectively. For the blue detuned case, $g_A =$ (-0.0016 $\pm$ 0.0002) nm$^{-1}$ based on the value of the gap from the atomic force micrograph and its propagated uncertainty. Using these values, parameters from our system, and Eq. (1), we find a predicted lasing threshold power of (1080 $\pm$ 23) µW, in good agreement with observations.

Oscillators can phase-lock to each other and to external inputs, which is commonly employed for injection locking optical lasers. In analogy to the optical case, we injection lock our metamolecule phonon laser (MPL) by using the electrostatic actuator to apply a weak mechanical force to the device. The large-amplitude laser output signal phase-locked to the applied RF tone, illustrating device utility for amplifying and detecting the frequency of weak frequency-modulated (FM) mechanical stimuli. A stimulus at varying input frequency $f_{in}$ was applied to the device operating a free-running frequency $f_0 \approx 3980$ kHz; the optical pump power was ≈ 2 mW. Figure 4a shows a log-scale contour plot of the power spectral density of the transduced motion for $f_{in}$ swept across $f_0$; linescans at various points are given in Fig. 4b. Within a band defined by $f_0 \pm f_L/2$, the MPL frequency follows $f_{in}$ synchronously, whereas outside this band, frequency pulling of $f_0$ toward $f_{in}$ is evident from the significant curvature in $f_0$, as are multiple distortion sidebands. These spectral signatures (locking, pulling, and sidebands) are theoretically described via

$$f_p = f_{in} + (p+1)\delta f \sqrt{1 - \left(\frac{f_L/2}{\delta f}\right)^2}, \qquad (2)$$

where $\delta f \equiv f_{in} - f_0$ is the detuning of the injection signal, $p$ is the sideband order, and $f_L$ is the injection locking range [25–29]. Interestingly, sonification of our MPL data clearly reveals the spectral signatures predicted by Eq. (2), including a clearly audible onset of locking (Supplementary Fig. 6, Supplementary Audio 1). As such, this technique may prove useful for future analyses of regenerative oscillators [30].

The locking range is governed by the ratio of the injected amplitude to the amplitude of the free running laser. In our experiments, the simple linear form of the range is given by

$$f_L = \frac{\omega_m}{2\pi} \frac{1}{Q_m} \frac{x_{inj}}{x_{free}} = \frac{\omega_m}{2\pi} \frac{x_{in}}{x_{free}}, \qquad (3)$$

where $x_{free}$ is the metamolecule displacement of the free-running MPL. The commonly used [25] injected displacement near resonance $x_{inj}$ is re-expressed through the amplitude $x_{in} = x_{inj}/Q_m$, which is the experimentally observable off-resonant displacement response to the drive force stimulus applied far below the resonance frequency. Dashed blue lines in Fig. 4a correspond to Eq. (2), wherein $f_0$ and $f_L$ are used as fitting parameters with values of 3980 kHz and 27 kHz, respectively. While the major features of Fig. 4a are captured using Eq. (2), there are significant deviations from this linear theory. The total observed range (36.9 kHz) at this injection amplitude extends beyond the range expected from the fit of Eq. (2) (Fig. 4c); this trend was observed for all $x_{in}$ applied. The extended locking range is accompanied by the appearance of strong sidebands (green curve, Fig. 4b) indicative of oscillations that likely result from nonlinear interaction between the drive amplitude and the free-running MPL. The amplitude and number of these sidebands increases significantly with increased injection amplitude (Supplementary Fig. 7). Furthermore, at large injection amplitudes the total observed locking range was smaller than that predicted by Eq. (3), as determined from calibrated displacement data for $x_{in}/x_{free}$. However, the measured range asymptotically approaches the linear prediction Eq. (3) in the low $x_{in}$ limit (Fig. 4d), indicating that linear injection locking theory is a valid quantitative description for our system at weak injection.

The injection locking results show that input displacements as small as 10 pm can be amplified by a large factor of about $x_{free}/x_{in}$ ($\approx 100$) over a bandwidth (the locking range) of 20 kHz to 50 kHz, an order of magnitude larger than the native mechanical linewidth of approximately 2 kHz. This is useful for amplifying and reading out frequency of weak FM tone signals, such as for tracking variations in frequency of another, coupled nanomechanical oscillator whose amplitude is otherwise too weak to detect directly. Notably, the MPL locking may be instrumental for reading out a different mechanical mode of the same nanomechanical system, including nonlinear coupled modes at integer multiples of the oscillator fundamental frequency [31]. The MPL can thus perform the function of a mechanical amplifier and phase-locked loop for frequency detection of various weak processes occurring at the nanoscale [32].

We have demonstrated an electrically-tunable nanoscale system that strongly couples the tightly-confined optical-frequency resonance of an individual plasmonic metamolecule with the system's mechanical and thermal modes. The metamolecule location engineers these interactions, distinguishing specific modes with spatial resolution given by the metamolecule size and, in principle, not limited by optical diffraction. Our active nanosystem allows facile electromechanical tuning of optical properties for strong electro-optic modulation. Introducing tailored coupling between mechanical and thermal modes enables electrical and optical excitation and control of rich nanomechanical dynamics, and shows regenerative amplification of weak mechanical stimuli in the RF frequency range by frequency injection locking. Nanoscale systems with such custom-engineered multi-domain interactions may find applications in nanomechanical sensing and signal transduction and form building blocks for future smart tunable metamaterials with unusual, coupled optical, mechanical and thermal properties.

Methods

*Fabrication processing:* Devices were produced using a combination of electron beam lithography, metal evaporation and liftoff, and reactive ion etching for patterning structural elements. Cantilevers are released and made free to vibrate using a combination of sacrificial layer, wet etching, and critical point drying. Full details of the fabrication process are given in the Supplementary Information.

*Uncertainties*: All uncertainties reported in this work are one standard deviation, derived or propagated from the corresponding fit parameter variances.

*Computational modelling:* We use a commercial finite-element solver for modelling electromagnetic, mechanical, and thermal behaviors of our devices. See Supplementary Information.

*Confocal spectroscopy:* To measure changes in the metamolecule optical resonance, a confocal spectroscopy setup based on a supercontinuum laser setup was used. A voltage was applied to devices and the recorded reflectance spectra were integrated for 10 ms, 0.5 s after the voltage is applied, after which the voltage is reset to zero to avoid slow drift and potential device damage due to surface charging of the dielectric. Reflectance data were fit to a Lorentzian curve, and from this procedure the wavelength $\lambda_{\text{LGP}}$ and bandwidth $\Delta\lambda$ of the metamolecule response were extracted. Uncertainties in Fig. 2d derived from the fit are much less than 1 % for $\lambda_{\text{LGP}}$.

*Driven motion detection*: The motion spectra of the cantilevers were measured by focusing a probe laser that is detuned from the metamolecule resonance wavelength, which generates a reflectance that is amplitude modulated at the mechanical frequency. In this experiment, devices operated in ambient conditions while a vector network analyzer supplied a small actuation voltage of ≈ 0.04 V with varying input frequency from 0.3 MHz to 100 MHz. Owing to an intrinsic DC bias, devices respond at this input frequency. The wavelength of the tunable probe laser was adjusted to varying detuning values, and the change in gap size resulting from electric actuation is mapped directly to changes in the reflected power as detected by the fast photodiode.

*Intrinsic DC bias*: The time-varying part of the electrostatic force applied to the cantilever from the actuators is $F_{\text{app}} \propto 2V_{\text{DC}}V_{\text{AC}} \sin \omega t + 1/2\, V_{\text{AC}}^2 \sin 2\omega t$, where $\omega$ is the frequency and $V_{\text{AC}}$ is specified voltage delivered. In our experiments, we set the offset voltage $V_{\text{DC,app}}$ to zero. However, the fact that we observe device response at both $\omega$ and $2\omega$ implies that there exists an intrinsic bias voltage in our devices $V_{\text{DC}} \equiv V_{\text{DC,app}} + V_{\text{DC,in}} = V_{\text{DC,in}}$. We determined $V_{\text{DC,in}}$ by measuring the relative output voltage, as a function of input $V_{\text{AC}}$, on an electronic spectrum analyzer at the input frequency and second harmonic, fitting these to linear and parabolic models, respectively, and using these fit coefficients along with the expression for $F_{\text{app}}$ to determine the relative weight of $V_{\text{DC}}$ and $V_{\text{AC}}$. From this procedure, we estimate an intrinsic bias of (+0.245 $\pm$ 0.013) V, where uncertainty is derived from the voltage fits.   The intrinsic DC bias is likely the result of trapped charges in the nitride from the electron-beam fabrication processing.

*Calibration of thermal motion data:* Calibration of motion data was performed using the equipartition method. See Supplementary Information for details.

**References**

1.  S. Jahani and Z. Jacob, "All-dielectric metamaterials," Nat. Nanotechnol. **11**, 23–36 (2016).

2.  N. J. Halas, S. Lal, W.-S. Chang, S. Link, and P. Nordlander, "Plasmons in strongly coupled metallic nanostructures," Chem. Rev. **111**, 3913–3961 (2011).

3.  M. Kadic, T. Bückmann, R. Schittny, and M. Wegener, "Metamaterials beyond electromagnetism.," Rep. Prog. Phys. **76**, 126501 (2013).

4.  T. Xu, A. Agrawal, M. Abashin, K. J. Chau, and H. J. Lezec, "All-angle negative refraction and active flat lensing of ultraviolet light.," Nature **497**, 470–474 (2013).

5.  C. Coulais, D. Sounas, and A. Alù, "Static non-reciprocity in mechanical metamaterials," Nature **542**, 461–464 (2017).

6.  K. L. Ekinci and M. L. Roukes, "Nanoelectromechanical systems," Rev. Sci. Instrum. **76**, 61101 (2005).

7.  M. Aspelmeyer, T. J. Kippenberg, and F. Marquardt, "Cavity optomechanics," Rev. Mod. Phys. **86**, 1391–1452 (2014).

8.  R. Thijssen, T. J. Kippenberg, A. Polman, and E. Verhagen, "Plasmomechanical resonators based on dimer nanoantennas," Nano Lett. **15**, 3971–3976 (2015).

9.  B. J. Roxworthy and V. A. Aksyuk, "Nanomechanical motion transduction with a scalable localized gap plasmon architecture," Nat. Commun. **7**, 13746 (2016).

10. N. I. Zheludev and E. Plum, "Reconfigurable nanomechanical photonic metamaterials," Nat. Nanotechnol. **11**, 16–22 (2016).

11. J.-Y. Ou, E. Plum, J. Zhang, and N. I. Zheludev, "An electromechanically reconfigurable plasmonic metamaterial operating in the near-infrared.," Nat. Nanotechnol. **8**, 252–5 (2013).

12. H. Zhu, F. Yi, and E. Cubukcu, "Plasmonic metamaterial absorber for broadband manipulation of mechanical resonances," Nat. Photonics **10**, 709–714 (2016).

13. D. F. P. Pile, T. Ogawa, D. K. Gramotnev, Y. Matsuzaki, K. C. Vernon, K. Yamaguchi, T. Okamoto, M. Haraguchi, and M. Fukui, "Two-dimensionally localized modes of a nanoscale gap plasmon waveguide," Appl. Phys. Lett. **87**, 261114 (2005).

14. M. Li, W. H. P. Pernice, and H. X. Tang, "Reactive cavity optical force on microdisk-Coupled nanomechanical beam waveguides," Phys. Rev. Lett. **103**, 223901 (2009).

15. T.-Z. Shen, S.-H. Hong, and J.-K. Song, "Electro-optical switching of graphene oxide liquid crystals with an extremely large Kerr coefficient.," Nat. Mater. **13**, 394–399 (2014).

16. B. S. Dennis, M. I. Haftel, D. A. Czaplewski, D. Lopez, G. Blumberg, and V. A. Aksyuk, "Compact nanomechanical plasmonic phase modulators," Nat. Photonics **9**, 267–273 (2015).

17. B. D. Hauer, C. Doolin, K. S. D. Beach, and J. P. Davis, "A general procedure for thermomechanical calibration of nano/micro-mechanical resonators," Ann. Phys. **339**, 181–207 (2013).

18. C. Metzger and K. Karrai, "Cavity cooling of a microlever," Nature **432**, 1002–1005 (2004).

19. J. B. Khurgin, M. W. Pruessner, T. H. Stievater, and W. S. Rabinovich, "Laser-rate-equation description of optomechanical oscillators," Phys. Rev. Lett. **108**, 223904 (2012).

20. M. Zalalutdinov, A. Zehnder, A. Olkhovets, S. Turner, L. Sekaric, B. Ilic, D. Czaplewski, J. M. Parpia, and H. G. Craighead, "Autoparametric optical drive for micromechanical oscillators," Appl. Phys. Lett. **79**, 695–697 (2001).

21. I. Bargatin and M. L. Roukes, "Nanomechanical analog of a laser: amplification of mechanical oscillations by stimulated zeeman transitions," Phys. Rev. Lett. **91**, 138302 (2003).

22. M. Bagheri, M. Poot, L. Fan, F. Marquardt, and H. X. Tang, "Photonic cavity synchronization of nanomechanical oscillators," Phys. Rev. Lett. **111**, 213902 (2013).

23. K. Vahala, M. Herrmann, S. Knünz, V. Batteiger, G. Saathoff, T. W. Hänsch, and T. Udem, "A phonon laser," Nat. Phys. **5**, 682–686 (2009).

24. H. Rong, R. Jones, A. Liu, O. Cohen, D. Hak, A. Fang, and M. Paniccia, "A continuous-wave Raman silicon laser," Nature **433**, 725–728 (2005).

25. R. Adler, "A study of locking phenomena in oscillators," Proc. IRE **34**, 351–357 (1946).

26. S. Knünz, M. Herrmann, V. Batteiger, G. Saathoff, T. W. Hänsch, K. Vahala, and T. Udem, "Injection locking of a trapped-ion phonon laser," Phys. Rev. Lett. **105**, 13004 (2010).

27. M. Wang, C. Zhao, X. Miao, Y. Zhao, J. Rufo, Y. J. Liu, T. J. Huang, and Y. Zheng, "Plasmofluidics: Merging Light and Fluids at the Micro-/Nanoscale," Small **11**, 4423–4444 (2015).

28. Y. Y. Liu, J. Stehlik, M. J. Gullans, J. M. Taylor, and J. R. Petta, "Injection locking of a semiconductor double-quantum-dot micromaser," Phys. Rev. A **92**, 53802 (2015).

29. A. E. Siegman, *Lasers* (University Science Books, 1986).

30. W. L. Diaz-Merced, R. M. Candey, N. Brickhouse, M. Schneps, J. C. Mannone, S. Brewster, and K. Kolenberg, "Sonification of astronomical data," Proc. Int. Astron. Union **285**, 133–136 (2011).

31. D. Antonio, D. H. Zanette, and D. López, "Frequency stabilization in nonlinear micromechanical oscillators," Nat. Commun. **3**, 806 (2012).

32. L. G. Villanueva, R. B. Karabalin, M. H. Matheny, E. Kenig, M. C. Cross, and M. L. Roukes, "A Nanoscale Parametric Feedback Oscillator," Nano Lett. **11**, 5054–5059 (2011).

**Acknowledgements**

We thank R. Ilic and I. Brener for a meticulous reading of the manuscript and fruitful discussions.

**Author Contributions**

B.J.R developed the fabrication process, designed and fabricated the devices, designed and performed the experiments, developed the computational models, analyzed the data, and wrote the manuscript. V.A.A developed the fabrication process, designed the experiments, developed the computational models, and wrote the manuscript.

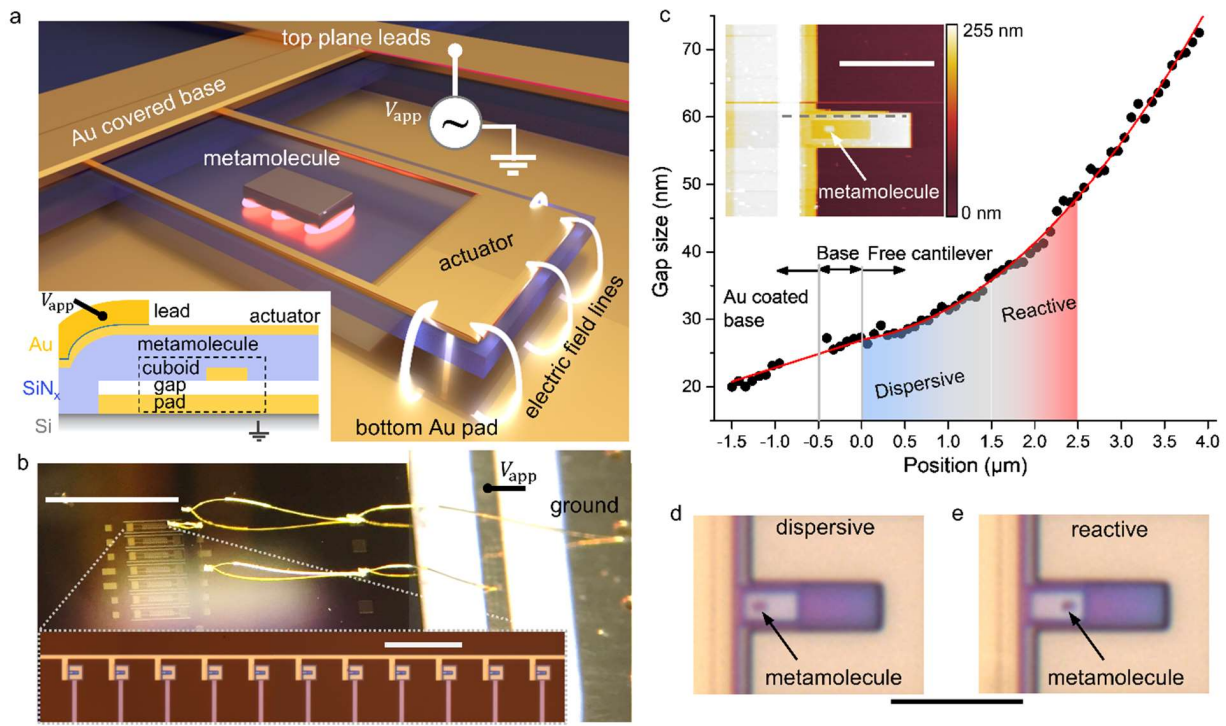

Figure 1. Tunable metamolecule nanosystem. (a) Illustration of the device architecture. The inset shows a side view of the device structural layers and elements. (b) Optical micrograph of fabricated chip packaged on a printed circuit board and electrically bonded; the inset shows a zoom-in of an array of devices. The scale bar is 1 mm (inset, 50 µm). (c) Line scan of an atomic force micrograph (gray dashed line, inset) of a 4 µm cantilever with an integrated metamolecule on the same chip as the devices studied in this work. The total gap size is the sum of the nominal 20 nm sacrificial layer thickness and the deflection due to residual stress; the profile is corrected for the ≈ 130 nm lead thickness. Red lines are fits that indicate the upward deflection of the structure due to tilt at the gold coated base (linear) and the curved shape of the free cantilever (parabolic). Regions of dispersive and reactive optomechanical coupling, based on simulation data, are indicated by the shaded region. Scale bar (inset) is 4 µm. (d) A dispersive, nominally 6 µm device having a metamolecule at ≈ 0.5 µm from the free base. (e) A comparable reactive device with ≈ 2.0 µm metamolecule position. Scale bar for d,e is 6 µm.

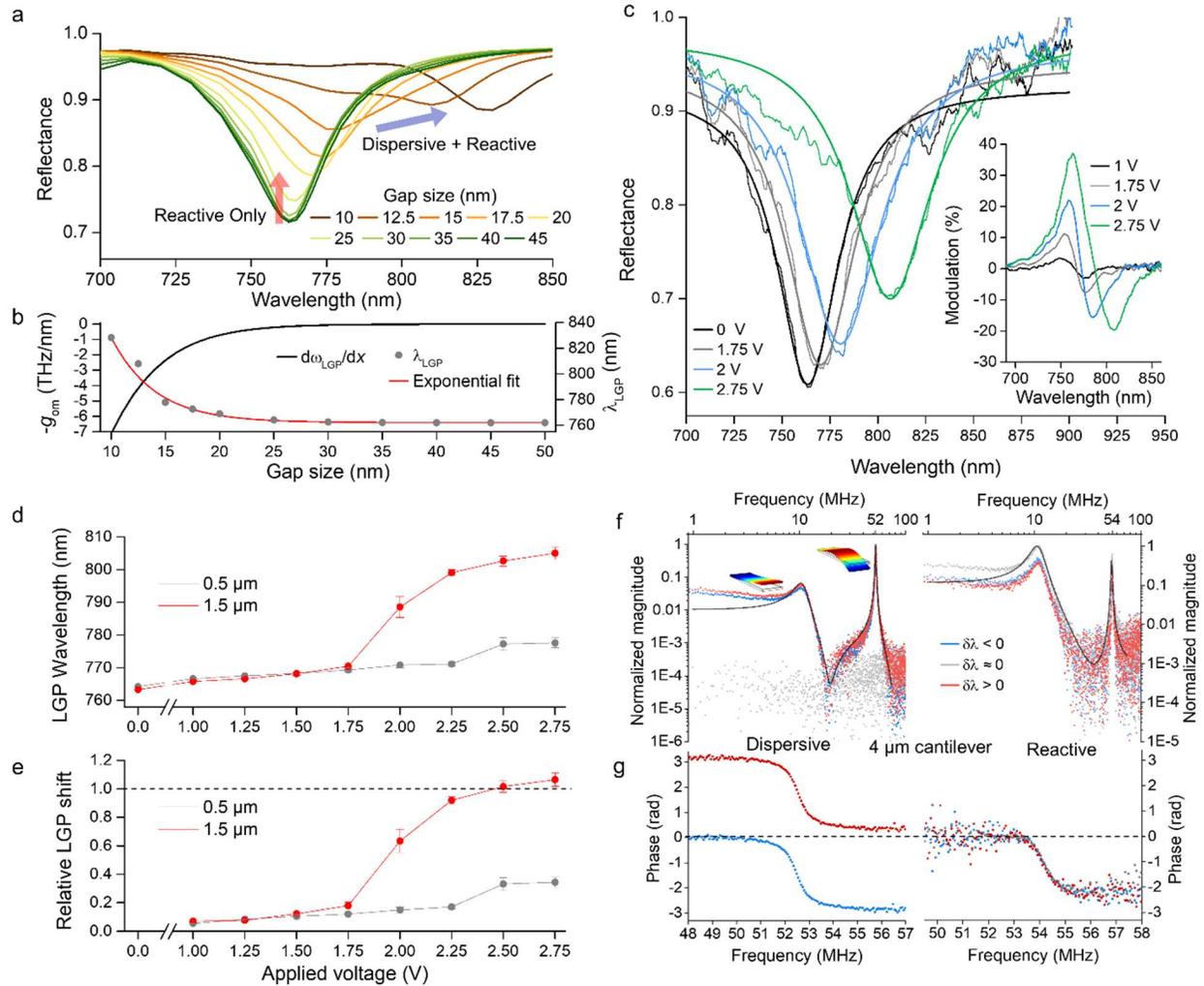

Figure 2. Electromechanical metamolecule tuning and optomechanical coupling measurement. Finite-element calculations of (a) reflectance of the LGP resonance as a function of gap and (b) extracted LGP wavelength (with exponential fit) and calculated dispersive optomechanical coupling constant. (C) Experimentally measured spectral reflectance of a single metamolecule as a function of applied voltage with Lorentzian fits; the inset shows the large amplitude modulation by the electrically actuated metamolecule. (d) Experimentally measured LGP wavelengths and (e) relative shift of the LGP resonance as a function of voltage; uncertainties are derived from Lorentzian fits. (f) Normalized magnitude of electrostatically driven motion for dispersive (left panels) and reactive (right panels) metamolecules measured with different probe laser detuning. Black lines correspond to a best fit of two Lorentzian functions added coherently; inset cartoons depict the shape of the flexural modes. The dispersive case illustrates that the metamolecule motion is described by a coherent superposition of the displacement from two flexural modes, the modal displacement magnitudes being comparable at that position on the cantilever. (g) Motion phase near the second order flexural mode resonance. Dispersive (reactive) coupling produces opposite (same) phase at different signs of detuning.

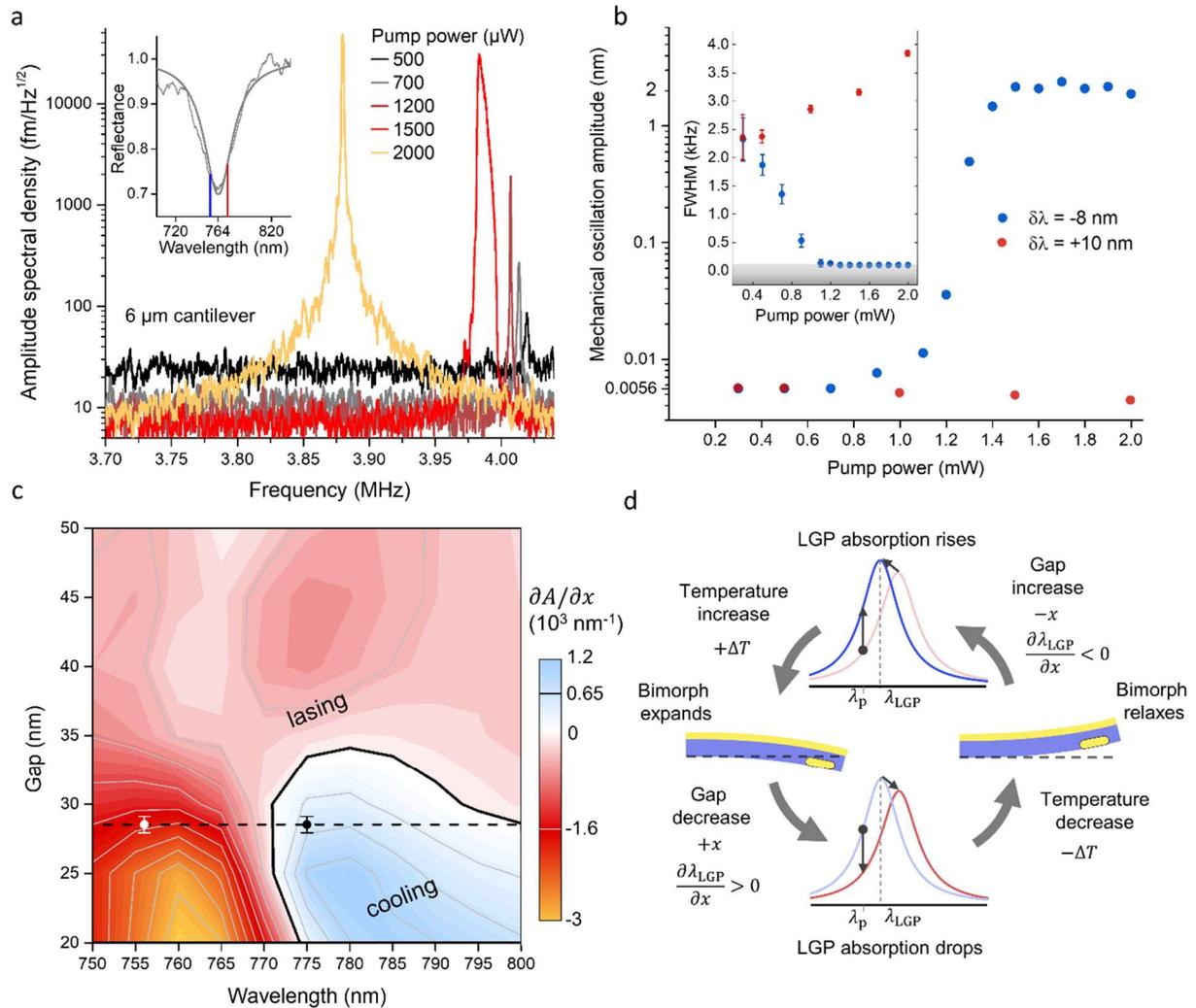

Figure 3. Metamolecule induced phonon lasing and cooling. (a) Mechanical amplitude spectral density for optical pump powers in different regimes of below threshold (black, gray), above threshold (dark red), and saturated (gold). There also exists an apparently chaotic regime [22] for pumping near saturation (red curve), whereby the amplitude and frequency fluctuate rapidly. In this case, the apparent broadened linewidth is a result of rapid frequency variation during the instrument averaging. The reduction in noise floor with increasing pump power is a result of decreased imprecision noise. The inset shows the reflectance spectrum of the metamolecule with negative and positive experimental wavelength detunings marked with blue and red lines, respectively. (b) Mechanical oscillation amplitude of the metamolecule on a logarithmic scale as function of pump power for negative (blue dots) and positive (red dots) wavelength detuning; inset shows corresponding mechanical linewidths and their uncertainties determined from Lorentzian fits. The shaded gray region indicates the measurement resolution limit (Methods). (c) Calculated optomechanical absorbance gain landscape (red: excitation, blue: cooling) with experimental measurement points for blue (white dot) and red (black dot) detuning; gap value and uncertainty is from the AFM measurement. The solid black line indicates the zero-gain contour and dotted line indicates the expected gap of the device. (d) Illustration of the metamolecule phonon laser cycle, in which time-delayed, thermally-induced backaction adds energy to the mechanical oscillations.

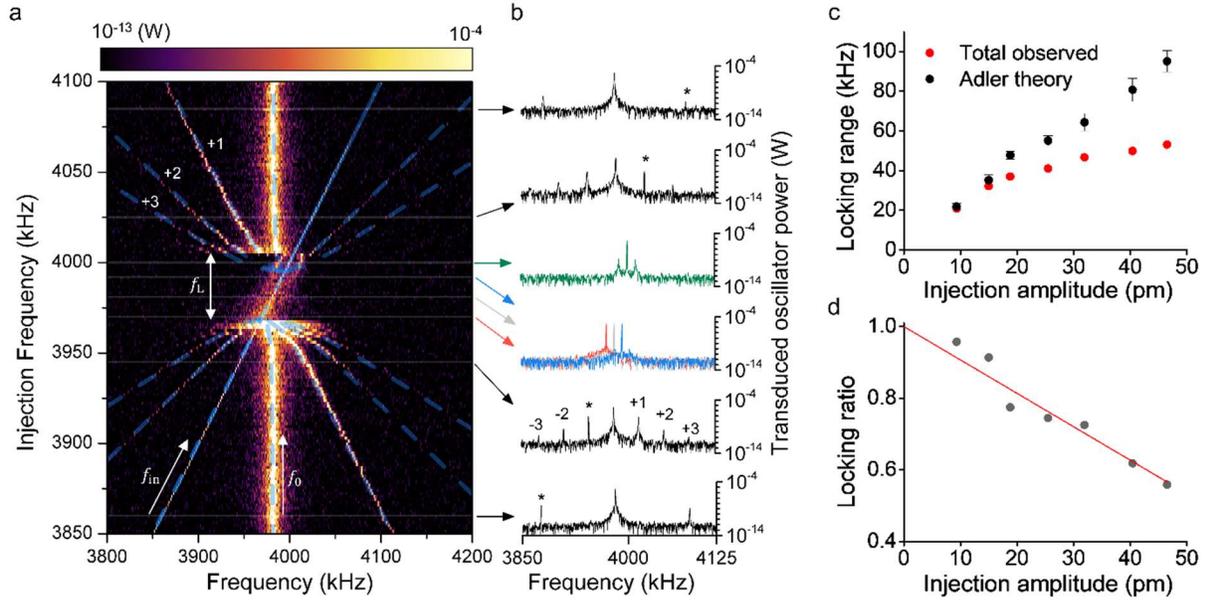

Figure 4. Injection locking of the metamolecule phonon laser. (a) Log-scale contour plot of the motion power spectral density of a metamolecule phonon laser (free-running frequency $f_0$) for varying injection frequency $f_{in}$ at an injection amplitude of $x_{in} \approx 25$ pm . The oscillator is locked to the input RF tone and follows $f_{in}$ over the interval $f_L$. (b) Line cuts of the motion power spectra for unlocked (black curves, $f_{in}$ marked by * symbol) and locked (red, gray, blue, and green curves) operation, plotted on log-scale. Numerical annotations represent the distortion sideband order up to a value of $\pm 3$. Additional sidebands appearing at the top of the locking range (green curve) are distinct from the distortion sidebands outside the locking range predicted by Eq. (2), and are likely the result of intermodulation between the applied actuated displacement and the oscillator. (c) Total observed locking range (red) (i.e., central rectangular band in (a)) and the Adler theory prediction $f_L = \omega_m/2\pi \, x_{in}/x_{free}$ (black); these quantities agree at low injection amplitude. The predicted values correspond to the average ratio of the injection displacement to the free-running displacement $x_{in}/x_{free}$ measured at far-detuned $f_{in}$, ranging from 3800 kHz to 3825 kHz; uncertainties are one standard deviation from multiple detunings. (d) Ratio of the total observed range to the theoretical predictions from Adler's theory, which shows a linear relationship converging to unity for weak injected signals.

## Supplementary Information

### Fabrication procedure

Devices are fabricated monolithically from a bare Si chip using repeated steps of aligned electron-beam lithography (aEBL), as depicted in Supplementary Fig. 1. Unless otherwise specified, each EBL exposure used a bilayer resist of 495k/950k molecular weight polymethyl methacrylate with total thickness three times greater than the total thickness of deposited metal layer. In the process, metal layers are formed using electron-beam evaporation and liftoff (LO) in a 1:1 by volume solution of acetone and methylene chloride. The first layer comprises a 5 nm Ti adhesion layer, 50 nm Au pad, and 20 nm Cr sacrificial layer. Cuboids with nominal dimensions of approximately $(350 \times 160 \times 40)$ nm$^3$ are then formed with an aligned EBL (aEBL) exposure. The nitride device layer is deposited at 180 °C using a plasma-enhanced chemical vapor deposition (PECVD) process incorporating inductively-coupled plasma (ICP) during the deposition. Stress in the nitride film is controlled via ICP power to fall within the range from 175 MPa to 225 MPa. Actuators, formed using a third aEBL+LO step, are composed of 15 nm Au atop a 3 nm Ti adhesion layer. Next, electrical leads and bond pads are formed from a 10 nm Ti, 120 nm Au stack using a fourth aEBL+LO step. A final aEBL step using a commercial high-resolution resist forms a dry-etch mask, which is transferred to the nitride using reactive-ion etching. The exposed Cr sacrificial layer is etched using a solution of ceric ammonium nitrate and chips are finalized by critical point drying in liquid $CO_2$. Devices are finalized by wire-bonding chips to a printed circuit board connectorized with coaxial RF jacks.

### Finite element modelling

We use a commercial finite-element solver to model electromagnetic, mechanical, and thermal properties of the devices. Two model geometries are used. The first (Supplementary Fig. 2a) is used for electromagnetic modelling and consists of a cylindrical domain comprising a 165 nm thick, 2 μm wide silicon nitride beam cantilever placed over an Au pad (atop a Si substrate) separated by a variable air gap. A 15 nm thick Au actuator is placed atop the beam with an approximately $(2.5 \times 1.75)$ μm$^2$ opening filled with air and centered above a cuboid with dimensions $(350 \times 160 \times 40)$ nm$^3$; a fillet with 15 nm radius of curvature is applied to all edges of the cuboid to avoid spurious electrical field effects. A Gaussian beam with focal point at the surface of the Au pad and spot radius of 0.6 $\lambda_0$/NA for numerical aperture NA = 0.3 or 0.9 and varying input wavelength $\lambda_0$ is introduced *via* a port boundary condition above the structure. The remaining boundary conditions include a second port at the bottom of the Si substrate, perfect magnetic conductors for symmetry, and perfectly matched layers along the top, bottom, and outer edges to prevent reflections into the computational domain. Reflectance is calculated using S-parameters $R = |S_{11}|^2$, where $S_{11}$ corresponds to the top port. Absorption is calculated using a loss-integral formulation, via

$$A \equiv \frac{1}{P_0} \int \frac{1}{2} \text{Re}[\boldsymbol{J} \cdot \boldsymbol{E}^*] dV, \tag{S1}$$

where $P_0$ is the power launched into the domain. The integral of the inner product of current density $\boldsymbol{J}$ and (complex conjugate) electric field $\boldsymbol{E}^*$ is evaluated over the volume $V$ comprising all gold surfaces; all other materials are assumed lossless and therefore do not contribute to the integral. Optical properties for Au and silicon nitride are taken from ellipsometric data of experimentally deposited films. Absorbance gain $g_A$ is calculated as the negative gradient of the absorbance landscape $-\nabla_x A(x, \lambda)$, wherein a lower numerical aperture of 0.3 NA is used to match the conditions of the lasing experiments.

The second geometry, used to compute mechanical eigenfrequencies, modal masses, static deflection and thermal responses of the devices, consists of a silicon nitride cantilever supporting a 15 nm thick Au actuator and attached to an outer frame having 1.5 μm extent (Supplementary Fig. 2b).

A 130 nm thick Au lead is placed overhanging a portion of the frame. Fixed boundary conditions are placed on the outer edges of the frame and the top edge temperature of the 130 nm lead is assumed to be 300 K. A cuboid measuring (350×160×40) nm³ is placed within the cantilever and dissipates a heat load of 50 µW, as derived from the electromagnetic power loss integral. The silicon nitride has density 2200 kg·m⁻³, elastic modulus 200 GPa, and Poisson ratio 0.2 [9]. Eigenmode calculations are performed to determine the vibrational frequencies of the cantilever, and the computed values agree with experimentally measured results to within 5 %. We evaluate the effective (modal) mass of the 6 µm cantilever by integrating the calculated mode shape $\Phi$ as

$$m_{\text{eff}} = \frac{1}{q^2} \int \rho(\mathbf{r}) \, |\Phi(\mathbf{r})|^2 \mathrm{d}V, \tag{S2}$$

where $q$ is a normalization factor representing a generalized coordinate for the cantilever displacement at the location of the embedded cuboid (the metamolecule), and $\rho(\mathbf{r})$ is position $\mathbf{r}$ dependent material density. A modal displacement of 0.1 times the maximum displacement at the tip as determined from finite element calculations. We find a modal mass of 2.11×10⁻¹⁵ kg for the fundamental mode.

The static downward deflection of the cantilever is produced by an elevated temperature of approximately 312 K, which is calculated as the average value over the free cantilever. The maximum downward displacement is normalized by the temperature increase of $\Delta T \approx$ 12 K, and the bimorph displacement gain $g_\text{B} \equiv \partial x / \partial \Delta T$ is evaluated at the cuboid location. To determine the bimorph displacement gain corresponding to the thermal excitation of the fundamental mechanical mode, we compute an overlap integral between the static displacement of the cantilever generated by the average temperature elevation of 12 K (50 pm K⁻¹ at the metamolecule location) and the modal displacement of the fundamental mode (Supplementary Fig. 2c), giving a final value of $g_\text{B} \approx$ 45 pm K⁻¹. Supplementary Fig. 2d shows the thermal time constant of the system, evaluated by applying the heat load as a step function at 0 s.

**Experimental setups**
Two experimental setups are used (Supplementary Fig. 3). In both experimental setups, devices are illuminated with laser light polarized along the cuboid long axis to excite the LGP mode that lies within a wavelength band from 750 nm to 820 nm. The first setup (Supplementary Fig. 3a) is a confocal optical spectroscopy microscope in which a supercontinuum laser spanning wavelengths from 600 nm to 1000 nm is focused onto the devices with a 0.9 NA objective. Devices operate in ambient air and are electrically connected to an alternating current voltage source, or the voltage output port of a vector network analyzer (VNA). For broadband measurements, reflected light from the supercontinuum laser is imaged onto an optical spectrometer and voltage is supplied from the voltage source. For motion measurements, the tunable laser is used, reflected light is imaged onto a fast photodiode (PD), and voltage is supplied from the network analyzer. The PD is connected directly to the VNA return port and a measurement of the $S_{21}$ parameter is performed.

In the second setup (Supplementary Fig. 3b), samples are placed in a vacuum chamber operating at a minimum pressure of ≈ 0.1 Pa (10⁻³ Torr) and electrically connected to an AC voltage source via an electrical feedthrough. Laser light from the tunable diode laser with a wavelength range of 756 nm to 800 nm is coupled into a single mode fiber and connected through a variable optical attenuator to a 50:50 fiber beamsplitter (tap), providing two outputs and one return path. One output of the tap is connected to an optical power meter, which is calibrated to measure the power delivered to the sample through the vacuum chamber cover glass by a fiber-coupled optical microscope. The microscope is connected to the second output and focuses a free-space beam, produced by a reflective achromatic collimator, onto the sample with a 0.3 NA objective. Reflected light is back-coupled into the fiber tap, wherein the return path focuses onto a PD connected to an electronic spectrum analyzer (ESA).

**Thermal motion calibration in vacuum**

The same transduction scheme is applied to devices placed in vacuum. In order to calibrate device thermal motion, the voltage power spectral density (PSD) is collected at low pump power ranging from 300 μW to 500 μW and fitted to Lorentzian curves, enabling extraction of the transduction gain in this linear region of device response [17]. We find transduction gains of (4.99 ± 0.87) mV nm$^{-1}$ at 300 μW with a linear slope of (0.035 ± 0.0012) mV nm$^{-1}$ μW$^{-1}$. Reported one standard deviation uncertainties of the transduction gain and its power-dependent slope are derived from parameter variances reported by the Lorentzian and linear fits, respectively. This linear increase in transduction gain with higher optical intensity results in a decrease of the input-referred detector dark noise and optical shot noise or equivalently, reduced imprecision noise in the motion readout. We apply linearly extrapolated calibration factors to lasing data over the full range of pump power. In order to extract the displacement, we utilize the equipartition theorem, whereby the integral of the displacement PSD corresponds to the RMS displacement of the cantilever at the location of the metamolecule, i.e., $1/2\pi \int S_{xx}(\omega)d\omega \approx x_{\text{eff}}^2$ [17,18]. Here, the effective displacement of the metamolecule is given by $x_{\text{eff}} = c_i x_{\text{rms}}$ with average displacement $x_{\text{rms}} = [k_B T_{\text{eff}}/(m_{\text{eff}}\omega_m^2)]^{1/2}$ where $k_B$ is Boltzmann's constant, $T_{\text{eff}}$ is the effective temperature, $m_{\text{eff}}$ is the modal mass, and $c_i \approx 0.1$ is the modal displacement coefficient of the metamolecule relative to the maximum tip-referenced value.

**Effective electro-optic Kerr coefficient**

The utility of our device architecture for electromechanically addressed optical modulation can be assessed by evaluating the effective electro-optic Kerr coefficient $\mathcal{K}$. Electrostatic force between two conductors with a potential difference is always attractive, and the direction of the plasmonic resonance shift is independent of the applied voltage polarity. Therefore the modulation at large voltage can be usefully described as Kerr effect with an effective $\mathcal{K}$. This parameter relates the change in optical phase $\Delta\varphi$ to the applied voltage through the following relation: $\mathcal{K} = \Delta\varphi \left(2\pi L_{\text{opt}} |E_{\Delta\varphi}|^2\right)^{-1} \approx d_V (4 V_\pi^2)^{-1}$, where $d_V$ is the effective electrical thickness of the cantilever with gap $g \approx 30$ nm, thickness $t_{\text{cant}} \approx$ 165 nm, $L_{\text{opt}} \approx 2\, d_V$ is the optical thickness of the modulator, and $V_\pi = E_\pi\, d_V$ is the applied voltage field required to induce a phase shift of $\pi$ rad [16]. Using experimentally measured parameters, we find $\mathcal{K} \approx$ 2 × 10$^{-8}$ m V$^{-2}$. This value is extremely large compared to natural Kerr materials [15].

**Motion transduction with smaller devices**

While the experiments in the main text focused on devices with metamolecules measuring approximately 350 nm in length and supporting $m = 3$ LGP modes, it is possible to achieve similar results with smaller metamolecules. Supplementary Fig. 5a shows scanning electron micrographs of two fabricated cantilevers measuring approximately 500 nm wide and 1.5 μm or 2 μm in length. Embedded within the cantilevers are cuboids measuring approximately (75×90) nm. The spectral reflectances of such metamolecules (Supplementary Fig. 5b) are characterized by $\lambda_{\text{LGP}} \approx 762$ nm and a quality factor of ≈ 10. Here, the lower quality factor strongly supports the observation of an $m = 1$ mode, whose quality is reduced due to increased radiative scattering. Supplementary Fig. 5c shows vector network analyzer measurements of the electrically stimulated cantilever motion in air, transduced by the smaller metamolecule. Clear vibrational peaks are visible at ≈ 44 MHz and 66 MHz for the 2 μm and 1.5 μm length devices, respectively, in good agreement with finite element calculations. Further, when the probe laser is translated slightly from the metamolecule position, the measured signal reduces to noise, indicating that transduction derives from the sub-diffraction sized metamolecule (see Supplementary Fig. 5).

**Control experiments**
We perform two experiments to ensure that measured motion signals and self-oscillation require the presence of the metamolecule. In the first we translate the probe laser from directly atop the metamolecule to locations ≈ 1 µm above (toward the cantilever tip) and below (toward the base). In both experimental configurations of vacuum with 0.3 NA (Supplementary Fig. 5a), and air with 0.9 NA (Supplementary Fig. 5b), we observe the detected signal dropping to that background noise level upon translation of the excitation beam. Vacuum measurements correspond to transduced thermal noise of the cantilever, whereas air measurements correspond to electromechanically driven measurements of the deveic. In the second experiment, we rotate the input polarization of the probe laser from parallel to the metamolecule long axis, which suppresses the LGP resonance (Supplementary Fig. 5c), and observe the signal of the metamolecule mechanical oscillator (MMO) dropping abruptly to noise for polarization angles beyond approximately 30° from parallel. Further, we find that for perpendicular polarization, we do not observe motion transduction at any pump power up to experimental limits of 2.75 mW. These data strongly indicate that the metamolecule is responsible for motion transduction and observed self-oscillation in this work [8,9].

**Photothermal threshold of the metamolecule phonon laser**
Self-oscillations in the cantilevers are driven by optical-frequency dependent absorption in the plasmonic resonator. The temperature increase due to absorption causes the bimorph to expand and deflect the cantilever downward from its equilibrium position, defined in the positive $x$-direction. An instability occurs, owing to the reduction in absorbed optical power with increasing $x$ resulting from the LGP resonance behavior which drives the cantilever into oscillation. The system is modeled using the following coupled equations

$$\frac{d\Delta T}{dt} = -\gamma_t \Delta T + \frac{1}{m_{th} c_p} P_{abs}(x), \tag{S3}$$

$$\frac{d^2 x}{dt^2} + \gamma_m \frac{dx}{dt} + \omega_m^2 x = \frac{1}{m_{eff}}[f_L + f_B(T) + f_{opt}], \tag{S4}$$

where $T$ is the device temperature, $\tau_t = \gamma_t^{-1}$ is the thermal time constant, $P_{abs}(x)$ is the displacement-dependent optical power absorbed in the device, and $\gamma_m = \omega_m/Q_m$ is the mechanical bandwidth of the cantilever with frequency $\omega_m$, quality factor $Q_m$. The thermal mass $m_{th}$ represents the full mass of the free cantilever and is ≈ 4.61×10$^{-15}$ kg. The device is driven by the Langevin force $f_L$ due to the thermal bath, the bimorph actuation force $f_B$, and the optical (radiation pressure) force $f_{opt} = g_{om}U/\omega_{LGP}$, where $U$ is average intracavity photon energy stored in the plasmonic resonator [7]

$$U = \left(1 - \sqrt{R_0}\right)\frac{(\gamma/2) P0}{\delta\omega^2 + (\gamma/2)^2}, \tag{S5}$$

where $R_0$ is the reflectance on resonance, $\gamma = \omega_{LGP}/Q_{LGP}$, $P_0$ is the optical power incident on the metamolecule, and $\delta\omega \approx \omega - \omega_{LGP}$ is the frequency detuning. Given that $f_B$ derives from optical absorption in the metamolecule, its magnitude is also proportional to $U$. However, using parameters determined below, we estimate the ratio $f_B/f_{opt} > 10^3$, indicating that radiation pressure is negligible compared to thermal backaction forces and can therefore be neglected.

To derive the threshold condition, we linearize the system around the equilibrium point $\bar{x}$ such that $x(t) = \bar{x} + x_0(t)$ for some small displacement $x_0$. In this limit, the change in the absorbed power and the bimorph actuation force $f_B$ scale linearly with $x_0$ and $\Delta T$

$$\Delta P_{\text{abs}} \equiv P_0 \frac{\partial A(\lambda,x)}{\partial x} x_0, \tag{S6}$$

$$f_B \equiv m_{\text{eff}} \omega_m^2 \frac{\partial x}{\partial \Delta T} \Delta T, \tag{S7}$$

where $A(\lambda, x)$ is the optical-wavelength and gap-size dependent absorbance of the LGP resonance, $\partial A/\partial x \equiv g_A$ is the absorbance gain, and $\partial x/\partial \Delta T \equiv g_B$ is the bimorph displacement gain. The system can then be recast as

$$\frac{d\Delta T}{dt} = -\gamma_t \Delta T + \frac{P_0}{m_{\text{th}} c_p} g_A x_0, \tag{S8}$$

$$\frac{d^2 x_0}{dt^2} + \gamma_m \frac{dx_0}{dt} + \omega_m^2 (x_0 - g_B \Delta T) = \frac{1}{m_{\text{eff}}} f_L, \tag{S9}$$

We note that the system described in Eq. (S8) and Eq. (S9) is the proper limit of other, more complete treatments that explicitly incorporate a third expression, describing the dynamics of the intracavity optical amplitude and coupling it to the mechanical coordinate [12]. Under the adiabatic approximation for optical amplitude, i.e. the assumption that the thermal time $\tau_t$ and mechanical frequency $\omega_m$ are much smaller than optical cavity loss rate $\gamma_{\text{LGP}} = \omega_{\text{LGP}}/Q_{\text{LGP}}$, the optical response follows instantaneously with the mechanical and thermal coordinates and there is negligible phase lag, therefore there is no contributions to excitation or damping from optical radiation pressure backaction. This simplification is well-justified in our experimental case, and indeed the results for the effective damping derived in our model are equivalent to those appearing elsewhere [12,31].

We proceed by Fourier transformation of Eq. (S8), from which we obtain the following expression for the temperature increase

$$\Delta T(\omega) = \frac{P_0}{m_{\text{th}} c_p} \frac{1}{(\gamma_t + i\omega)} g_A x_0, \tag{S10}$$

where $\omega$ is the Fourier frequency. Applying Fourier transformation to Eq. (S9), using Eq. (S10), and collecting imaginary terms, we find the effective damping coefficient

$$\gamma_{\text{eff}} = \gamma_m + \omega_m^2 \frac{P_0}{m_{\text{th}} c_p} \frac{g_A g_B}{\omega^2 + \gamma_t^2}, \tag{S11}$$

where the second term describes the contribution of the photothermal spring to the overall system damping. Threshold is crossed when $\gamma_{\text{eff}}$ passes through zero and becomes negative, which requires that $g_A < 0$, leading to the expression for threshold power at the mechanical resonance $\omega = \omega_m$

$$P_{\text{thresh}} = \frac{m_{\text{th}} c_p}{|g_A|} \frac{\omega_m^2 + \gamma_t^2}{Q_m \omega_m g_B}. \tag{S12}$$

As shown in Fig. 3a, the condition $g_A < 0$ is met both for blue-detuned dispersive devices as well as reactive devices with any detuning. This unique feature enables self-oscillation for a wide variety of device configurations.

**Frequency behavior of the metamolecule phonon laser**

For pump power just above threshold, the metamolecule phonon laser MPL frequency and amplitude are stable and characterized by a narrow linewidth (dark red curve, Fig. 3a of the main text). Similarly, for pump power above ≈ 1500 µW the laser frequency is stable and displays a narrow linewidth on top of a broad background (gold curve, Fig. 3a) with saturated amplitude. However, pumping near the high end of the transition region, but below saturation, produces apparently chaotic behavior characterized by rapid variations in frequency and amplitude, consistent with previous observations [22]. The red curve in Fig. 3a represents one such example of chaotic behavior, averaged with 100 Hz bandwidth. While the exact nature of this rapid frequency variation is not known, it is possible to measure a stable frequency within the resolution bandwidth filter time (FWHM data in Fig. 3b inset), and thus it is possible that the observed behavior derives from environmental perturbations to the setup.  The ≈ 0.25 % reduction in mechanical frequency observed below threshold does not originate from an optical spring, owing to the relatively low quality factor of the LGP resonance  [7].  Whereas this minute shift may be attributed to temperature-dependent reduction in the elastic modulus of the silicon nitride, the larger frequency reduction for high pump power is likely the result of an amplitude-dependent frequency nonlinearity occurring for strong pumping.

In measuring the linewidth of the MPL (Fig. 3b in the main text), drift and vibrations in our experimental system currently make long-term measurements problematic.  We therefore limit the resolution bandwidth of the ESA to 100 Hz to provide a compromise between resolution of the self-oscillation linewidth (higher resolution requiring longer sweep time) and elimination of systematic drift.  Nevertheless, this measurement setting is within an order of magnitude of the expected saturation value of the linewidth of ≈ 10 Hz  [19], and the observed narrowing (broadening) of the linewidth for blue-detuned (red-detuned) pumping is consistent with expectations for our system.  Furthermore, the uncertainty in the measurements is reasonably small.

**Sonification of lasing data**

We have performed sonification [28] of real-time injection locking data from one of our devices (Supplementary Fig. 6, Supplementary Audio 1). We used a ≈ 6 μm device with ≈ 19  pm (100 mV) injection amplitude.  The horizontal axis frequencies in Supplementary Fig. 5 were mapped to the audible range from approximately 200 Hz to 40000 Hz.  We provide a string of 188 segments taken from a continuous voltage trace during the locking experiment.  Each sequential segment is ≈ 0.25 s in length, spaced equidistantly, and normalized in amplitude.  The key features of the locking experiment are evident from the audio file. The initially free-running MPL is evident as the strong tone with frequency of ≈ 3 kHz at the beginning of the audio. The injected frequency, the MPL, and the generated sidebands are audible at an initial frequency of ≈ 200 Hz and increase in pitch until approximately 16 s into the audio. At this point, the signal abruptly changes to a nearly pure tone which increases in pitch with time, signifying locking. The data in Supplementary Fig. 5 contains a stitch between two contiguous frequency sweeps at an injection frequency of ≈ 4300 kHz, resulting from a spurious electronic signal, which produces an audible distortion near 22 s in the audio trace.

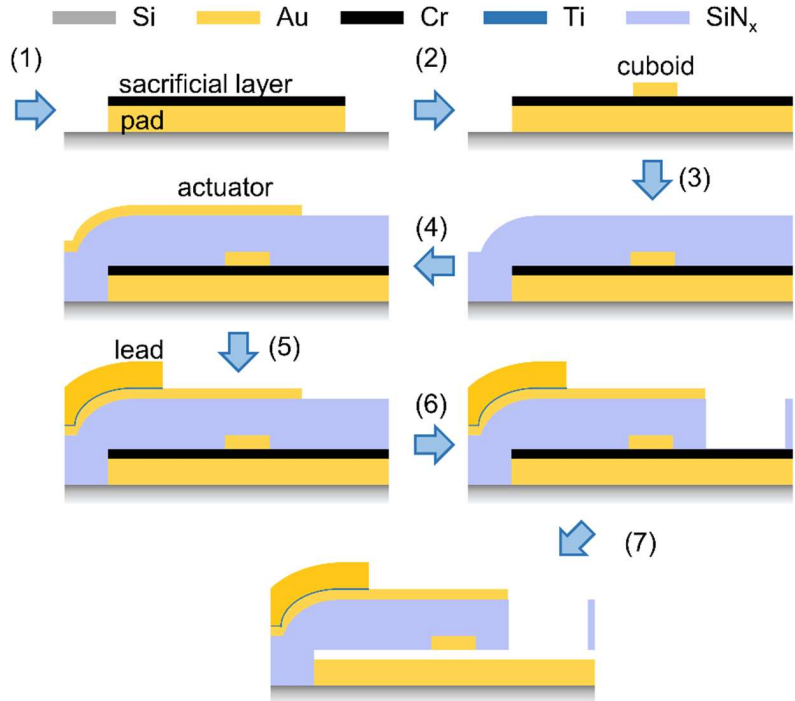

Supplementary Figure 1. Process flow for the tunable metamolecule nanosystem. Processes are: (a)EBL – (aligned) electron beam lithography, LO – liftoff, PECVD – plasma-enhanced chemical vapor deposition, RIE – reactive-ion etching, Etch – wet-chemical etching of Cr, CPD – critical point drying. The cutaway near the cuboid is included for illustration purposes and is not present in actual devices.

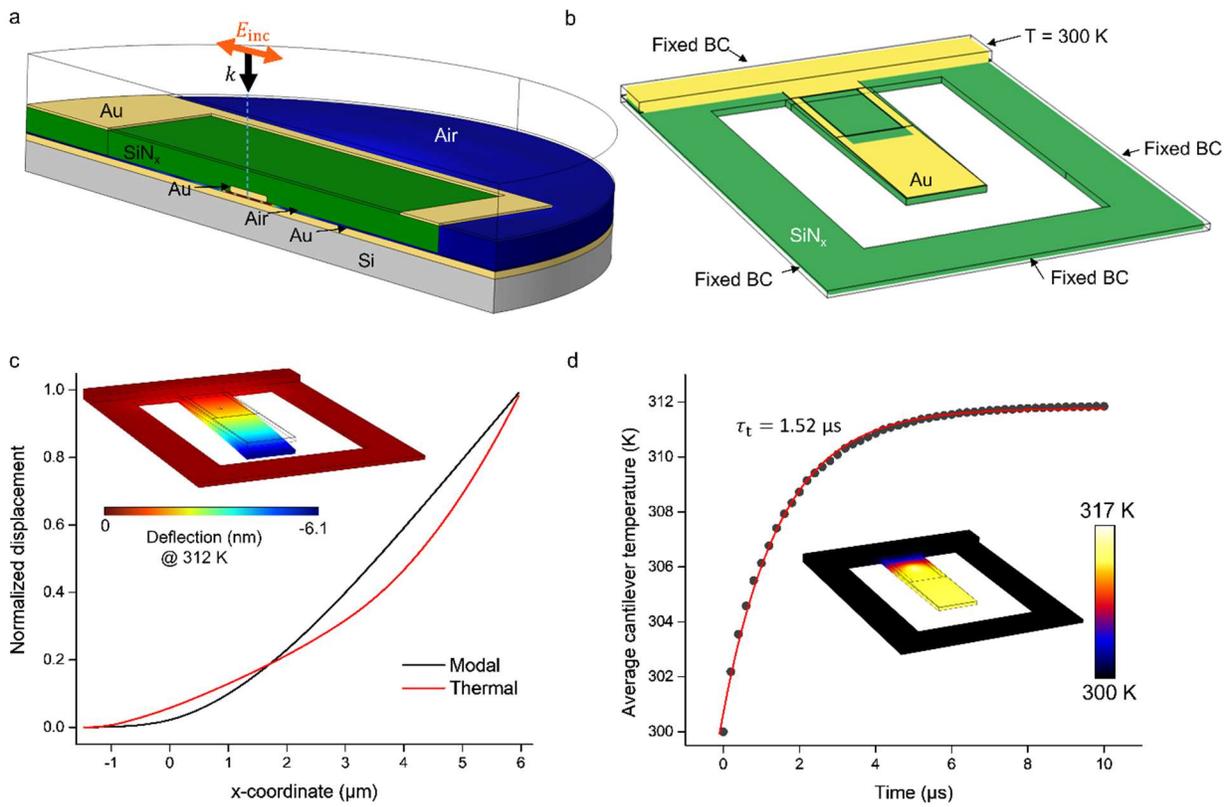

Supplementary Figure 2. Finite-element models and results. (a) Domain for electromagnetic calculations representing one half of a 4 μm length cantilever. The red arrow shows the incident electric field polarization ($E_{\text{inc}}$) whereas the black arrow shows the incident wavevector ($k$). (b) Domain for mechanical calculations. (c) Normalized absolute value of bimorph deflection (red) and modal deflection. The inset shows a 3D view of the total bimorph deflection (magnitude exaggerated) for a 50 μW heat load resulting in a 312 K average temperature. (d) Transient average temperature for the cantilever with fit to a first-order system response. Inset shows the temperature distribution of the cantilever after 10 μs of heating.

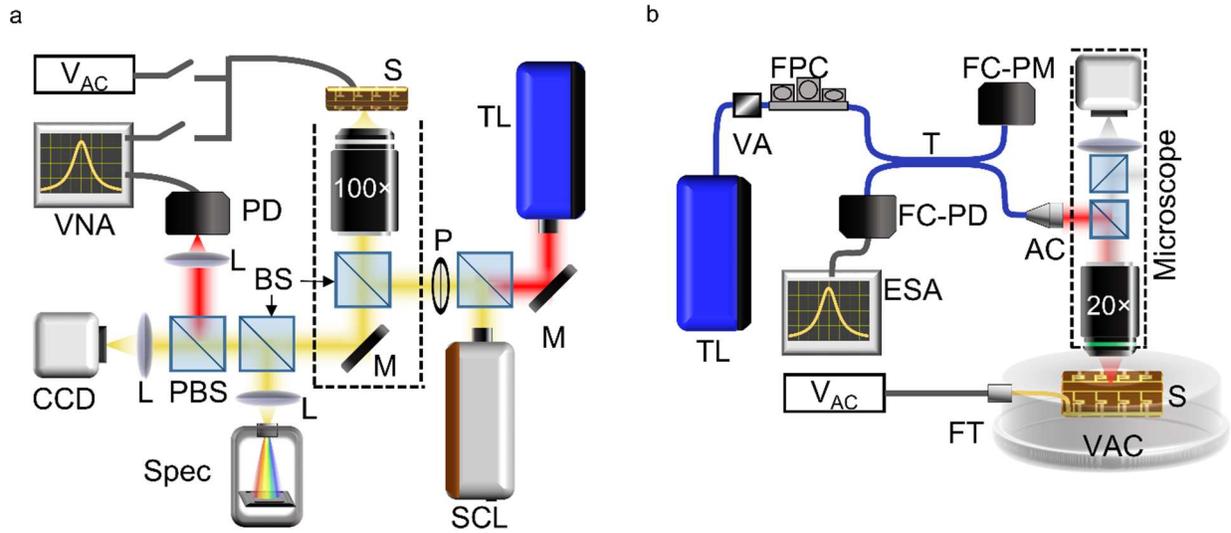

Supplementary Figure 3. Experimental setups. (a) Setup for spectroscopy and meta-molecule motion detection in ambient conditions. Components are: SCL – supercontinuum laser, TL – tunable diode laser; M – mirror, P – polarizer, BS – 50:50 beamsplitter, L – lens, Spec – spectrometer, PBS – polarizing 50:50 beamsplitter, PD – photodiode, VAC – alternating current voltage source, VNA – vector network analyzer, S – sample. (b) Setup for vacuum probing of devices. Components are: TL – tunable diode laser, VA – variable attenuator, FPC – fiber polarization controller, T – 50:50 fiber tap, FC-PM – fiber-coupled optical power meter, AC – achromatic collimator, VAC – vacuum chamber, FC-PD – fiber-coupled photodiode, FT – vacuum feedthrough. In both setups, components within dashed black lines constitute a microscope.

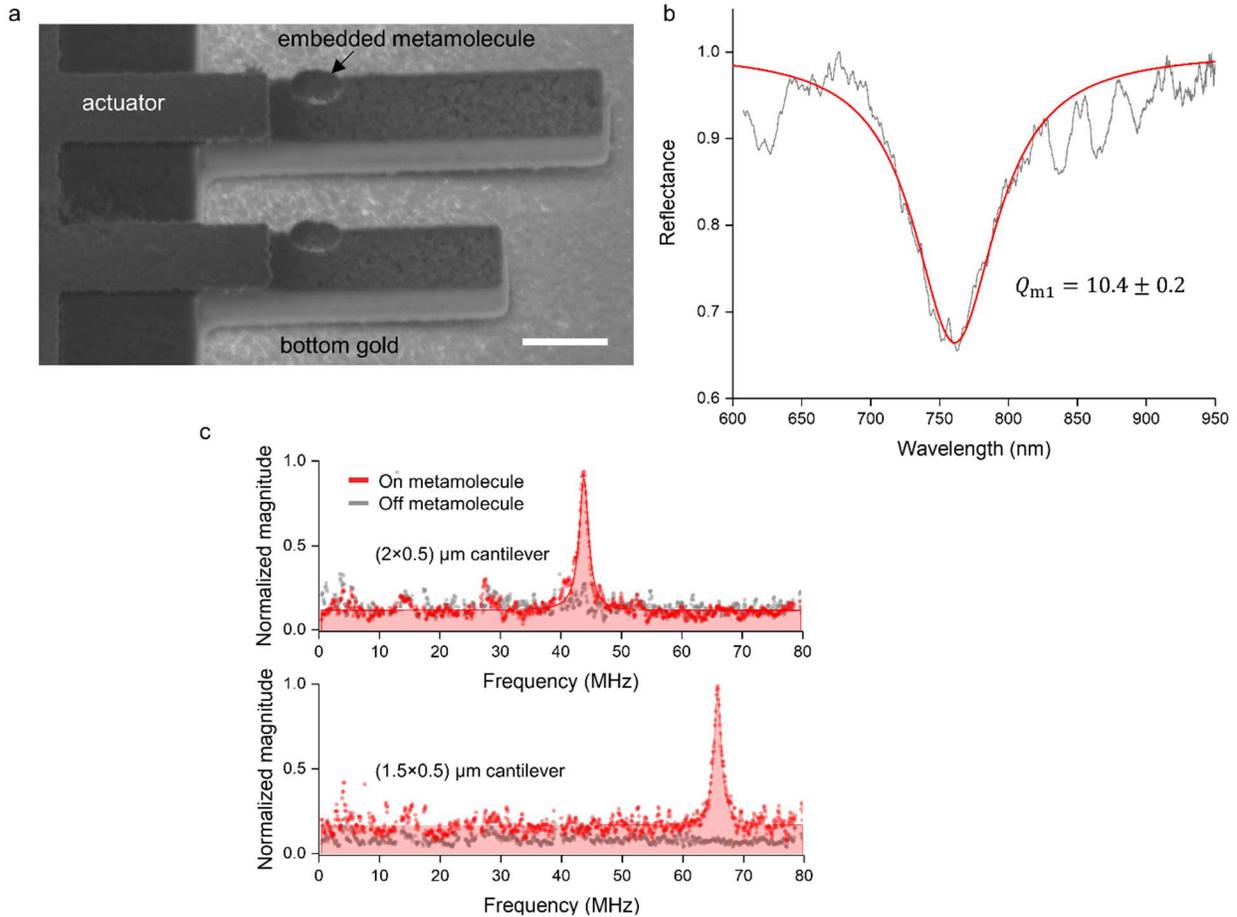

Supplementary Figure 4. Sub-diffraction metamolecule motion transduction. (a) Scanning electron micrographs of two fabricated cantilever devices measuring approximately 500 nm wide and 1.5 μm (bottom cantilever) or 2 μm (top) long. The ≈ (75×90×40) nm metamolecule, embedded on the underside of the cantilever, forms a visible 'bulb' through the thickness and protruding from the top of the silicon nitride [9]; scale bar is 500 nm. (b) Measured LGP reflectance spectrum revealing the resonance quality factor of $10.4 \pm 0.2$; the lower quality factor, compared to devices in the main text, is a characteristic of the $m = 1$ LGP mode. (c) Transduced motion spectra from the metamolecule for the 2 μm (top panel) and 1.5 μm (bottom panel) cantilevers, respectively.

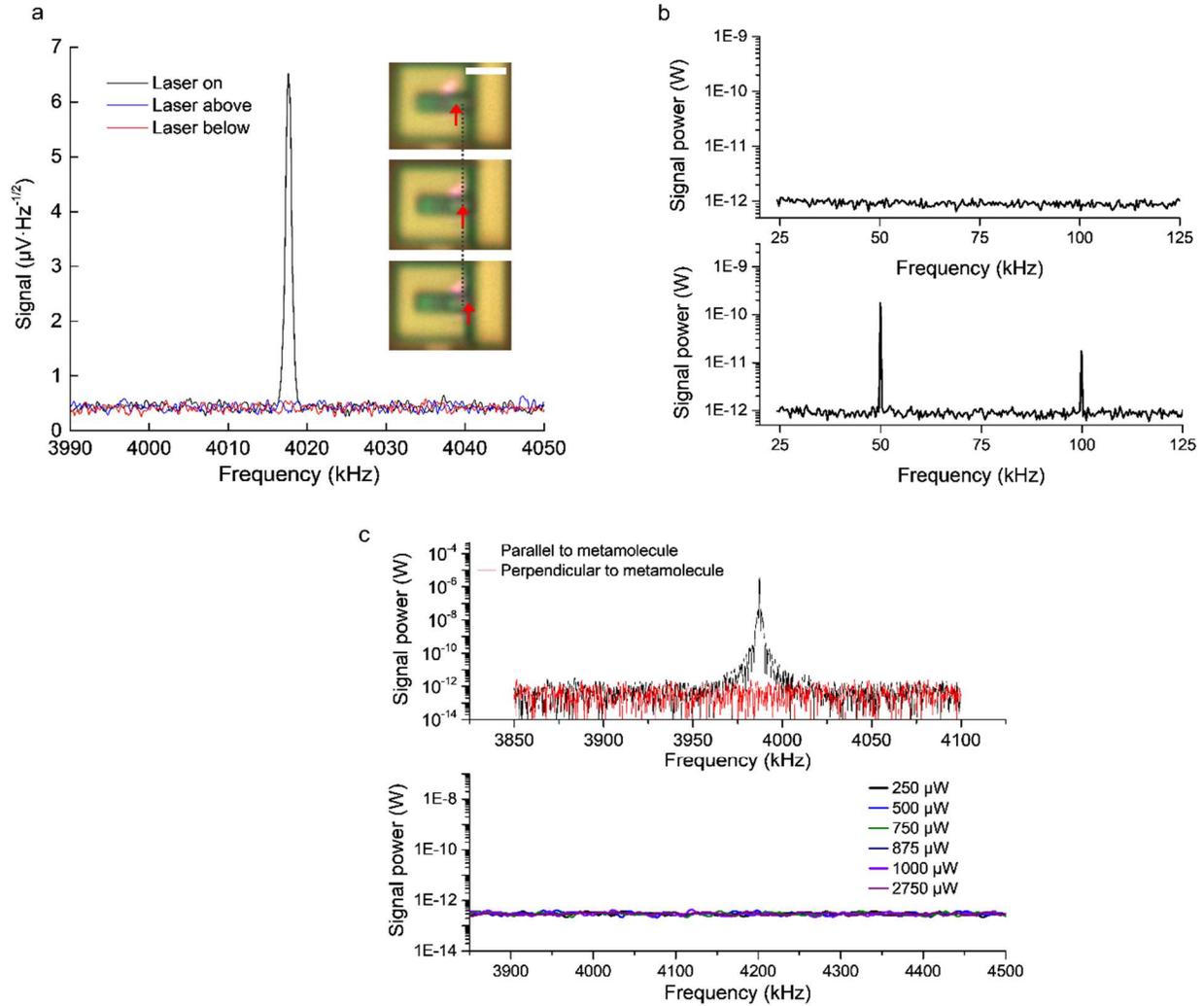

Supplementary Figure 5. Control experiments for metamolecule phonon lasing. (a) Position dependence of thermal motion transduction in the vacuum setup with 0.3 NA objective. Inset shows optical micrographs of the ≈ 6 μm cantilever under test, with the dotted black line marking the location of the metamolecule and the red arrow marking the focused beam centroid (located at approximately +1 μm, 0 um and −1 μm). The spurious light above the cantilever in the inset is due to multiple reflections in the imaging system. Scale bar is 5 μm. (b) Position dependence of driven motion detection in air (0.9 NA objective setup) with a 50 kHz drive frequency for laser positions (top panel) 1 μm above the metamolecule toward the cantilever tip and (bottom panel) atop the metamolecule. (c) Polarization dependence experiments showing (top panel) the MMO response with polarization parallel to the long axis of the cuboid comprising the metamolecule and perpendicular; the bottom panel shows data for perpendicular polarization as a function of pump power up to the system limit of 2.75 mW. Data in (c) correspond to the same device as those in (a).

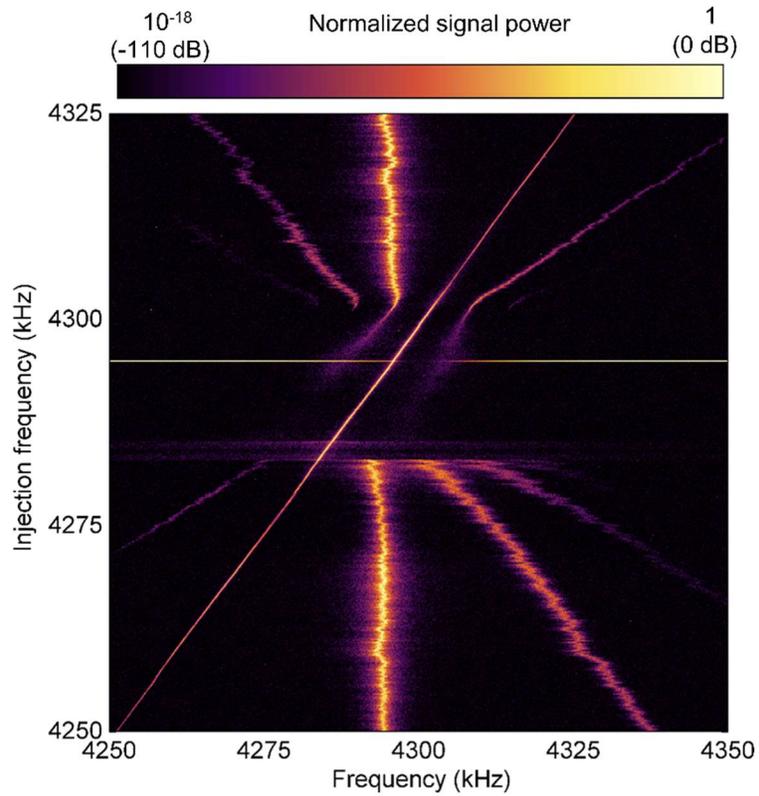

Supplementary Figure 6. Injection locking data for sonification. Contour plot of injection locking of a 6 μm cantilever device using 19 pm (100 mV) injection amplitude. The injection frequency sweep is from the bottom left ≈ 4250 kHz to the top right ≈ 4325 kHz, occurring over ≈ 48 s. The stitch in the data occurring at 4290 kHz is due to a spurious signal and is not representative of injection locking dynamics.

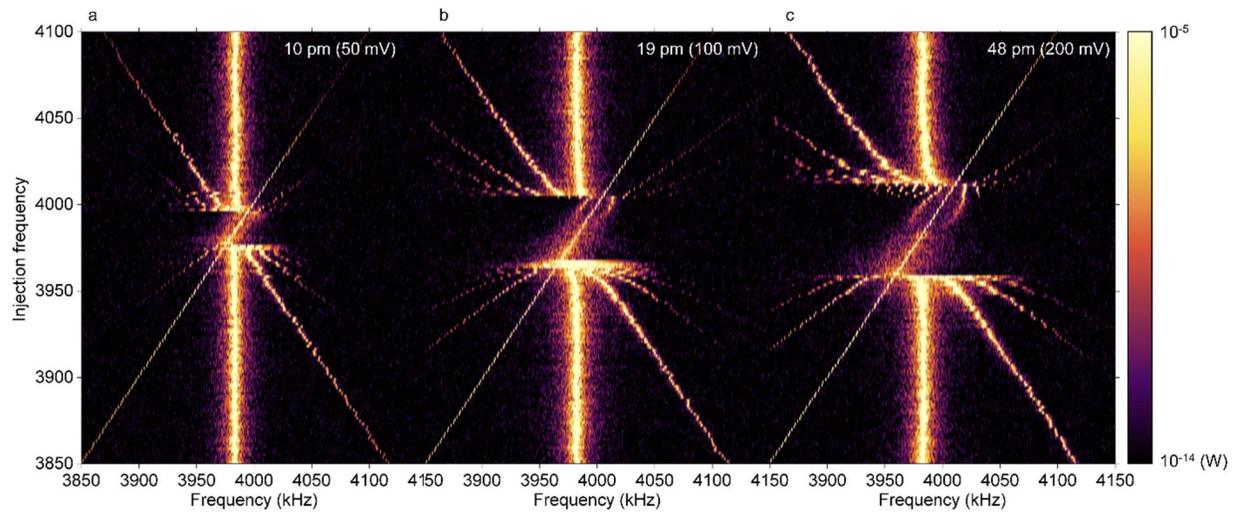

Supplementary Figure 7. Injection locking examples. A suite of injection locking contour plots for injection amplitudes (applied AC voltages) of approximately (a) 10 pm (50 mV), (b) 19 pm (100 mV), and (**c**) 48 pm (200 mV). The injection sweep frequency direction is from bottom left to top right. Nonlinear sidebands within the locking range are evident at increasing injection amplitude.

Supplementary Video 1. Electromechanical switching of metamolecule resonance. Time series at one-quarter speed of the metamolecule reflectance spectrum (gray circles) as a square-wave voltage with peak amplitude of 2.75 V and 4 Hz frequency is applied. The red lines are fits of the spectrum to Lorentzian curves.

Supplementary Audio 1. Sonification of metamolecule phonon laser data. Injection-locked MPL time-series converted to audio frequencies. Locking can be heard at ≈ 16 s as the abrupt transition to a nearly pure tone.

**References**

7.  M. Aspelmeyer, T. J. Kippenberg, and F. Marquardt, "Cavity optomechanics," Rev. Mod. Phys. **86**, 1391-1452 (2014).
8.  R. Thijssen, T. K. Kippenberg, A. Polman, and E. Verhagen, "Plasmomechanical resonators based on dimer nanoantennas," Nano Lett. **15**, 3971-3976 (2015).
9.  B. J. Roxworthy and V. A. Aksyuk, "Nanomechanical motion transduction with a scalable localized gap plasmon architecture," Nat. Commun. **7**, 13746 (2016).
12. H. Zhu, F. Yi, and E. Cubukcu, "Plasmonic metamaterial absorber for broadband manipulation of mechanical resonances," Nat. Photon. **10**, 709-714 (2016).
15. T.-Z. Shen, S. –H. Hong, and J. –K. Song, "Electro-optical switching of graphene oxide liquid crystals with an extremely large Kerr coefficient," Nat. Mater. **13**, 394-399 (2014).
16. B. S. Dennis, M. I. Haftel, D. A. Czaplewski, D. Lopez, G. Blumberg, and V. A. Aksyuk, "Compact Nanomechanical plasmonic phase modulators," Nat. Photonics **9**, 267-273 (2015).
17. B. D. Hauer, C. Doolin, K. D. S. Beach, and J. P. Davis, "A general procedure for thermomechanical calibration of nano/micro-mechanical resonators," Ann. Phys. **339**, 181-207 (2013).
18. C. Metzger and K. Karrai, "Cavity cooling of a microlever," Nature **432**, 1002-1005 (2004).
19. J. B. Khurgin, M. W. Pruessner, T. H. Stievater, and W. S. Rabinovich, "Laser rate-equation description of optomechanical oscillators," Phys. Rev. Lett. **108**, 223904 (2012).
22. M. Bagheri, M. Poot, L. Fan, F. Marquardt, and H. X. Tang, "Photonic cavity synchronization of nanomechanical oscillators," Phys. Rev. Lett. **111**, 213902 (2013).
30. W. L. Diaz-Merced, R. M. Candey, N. Brickhouse, M. Schneps, J. C. Mannone, S. Brewster, and K. Kolenberg, "Sonification of astronomical data," Proc. Int. Astron. Union **285**, 133-136 (2011).